\documentclass[floatfix,twocolumn,showpacs,prl,superscriptaddress]{revtex4}
\usepackage{graphicx}
\usepackage{dcolumn}
\usepackage{amsmath}

\makeatother

\begin{document}

\title{Impact of high-energy tails on granular gas properties} 
\author{Thorsten P\"oschel}
\affiliation{Charit\'e, Augustenburger Platz, 10439 Berlin, Germany}
\author{Nikolai V. Brilliantov}
\affiliation{Institute of Physics, University of Potsdam, Am Neuen Palais 10, 14469 Potsdam, Germany and \\
  Department of Physics, Moscow State University,  Vorobievy Gory 1, 119899 Moscow, Russia }
\author{Arno Formella} 
\affiliation{Universidad de Vigo, Department of Computer Science, Edificio Polit\'ecnico, 32004 Ourense, Spain}

\date{\today}

\begin{abstract} 
  The velocity distribution function of granular gases in the
  homogeneous cooling state as well as some heated granular gases
  decays for large velocities as $f\propto\exp(- {\rm const.} v)$.
  That is, its high-energy tail is overpopulated as compared with the
  Maxwell distribution. At the present time, there is no theory to
  describe the influence of the tail on the kinetic characteristics of
  granular gases. We develop an approach to quantify the overpopulated
  tail and analyze its impact on granular gas properties, in
  particular on the cooling coefficient.  We observe and explain
  anomalously slow relaxation of the velocity distribution function to
  its steady state.
\end{abstract}


\maketitle

\begin{figure}[htbp]
\centerline{\includegraphics[width=6.8cm,clip]{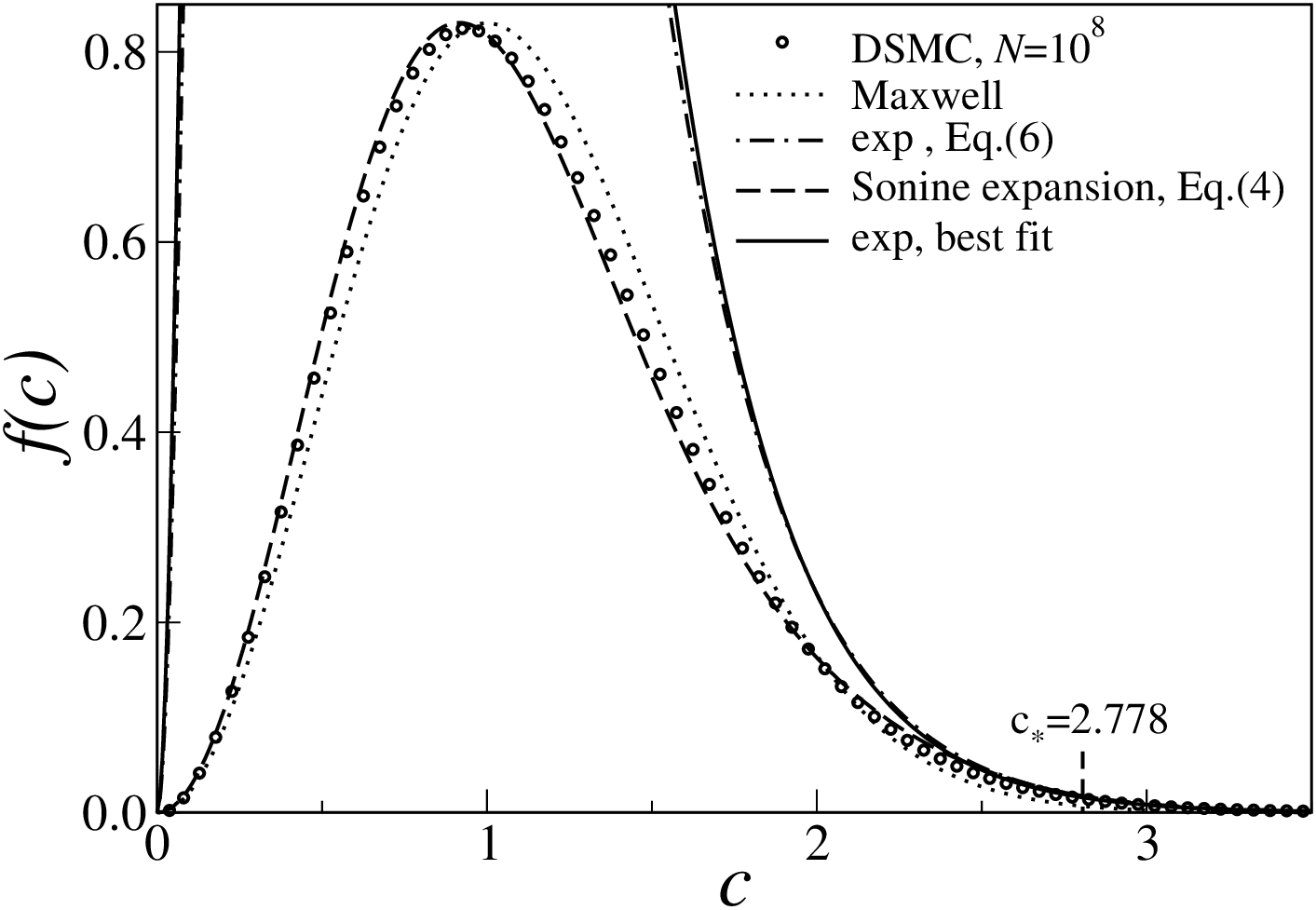}}
\centerline{\includegraphics[width=6.8cm,clip]{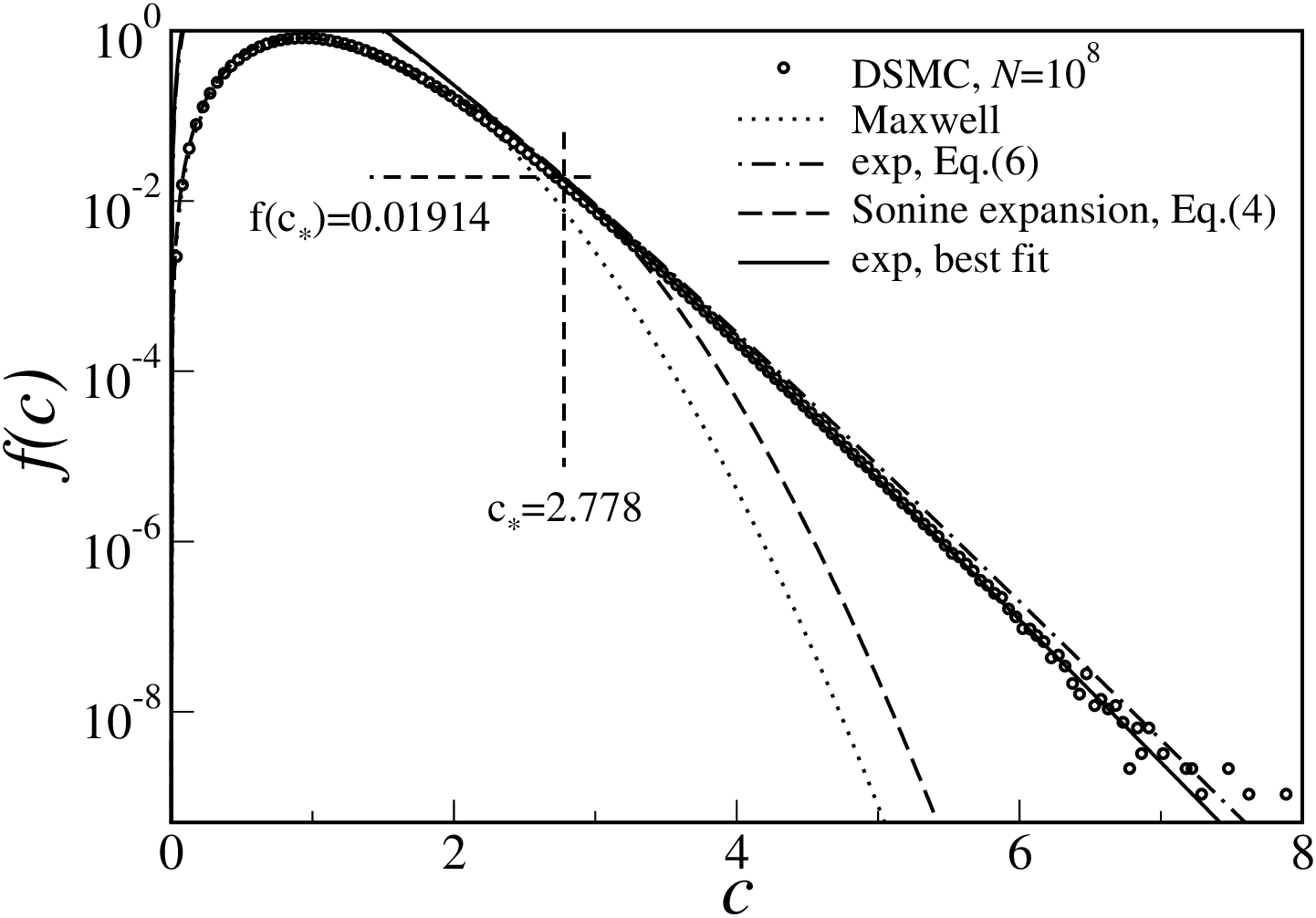}}
\caption{Velocity distribution function of a granular gas
  $\tilde{f}\left(\vec{c}\,\right)$ as normal and logarithmic plot.
  The symbols show a simulation of $N=10^8$ particles for
  $\varepsilon=0.3$. For $c\sim 1$, $\tilde{f}\left(\vec{c}\,\right)$
  is well described by the second order Sonine expansion, Eq.
  \eqref{eq:Sonexp} (top). For comparison the Maxwell distribution is
  also shown. For $c\gg 1$ the distribution function decays
  exponentially slow (bottom), see Eq. \eqref{eq:tail}.  The tail
  starts at $c\approx c_*$, Eq.~\eqref{eq:eq_for_cstar}.}
\label{fig:distribution}
\end{figure}

When a homogeneous granular gas evolves in the absence of external
forces, it develops a velocity distribution similar to a molecular
gas, however, its temperature decays due to the dissipative nature of
particle collisions. The velocity distribution function of granular
gases has attracted much scientific attention since it deviates
characteristically from the Maxwell distribution. There are two types
of deviations: First, there are deviations from the Maxwellian in the main
part of the distribution
\cite{GoldshteinShapiro1:1995,NoijeErnst:1998}, where the particle
velocities are close to the thermal velocity $v_T(t)= \sqrt{2T}$ (unit
particle mass is assumed). Second, the high-energy part,
$ v \gg v_T$,
deviates in its functional form,  i.e., the
distribution function decays as $f\propto\exp(-{\rm const.} v)$
\cite{EsipovPoeschel:1995,NoijeErnst:1998}, instead of
$f\propto\exp(-{\rm const.} v^2)$  as expected for 
the Maxwell distribution.  Heated granular gases with a Gaussian
thermostat are equivalent to gases in the homogeneous cooling state (HCS)
\cite{MontaneroSantosTailsGG:1999}.  That is,
the addressed properties apply for a wide class of granular gases. Figure
\ref{fig:distribution} shows the velocity distribution function
and its
deviation from the Maxwellian  (details see below).

Both types of deviations are characterized by the coefficient of restitution
$\varepsilon$ describing the post-collision particle  velocities
$\vec{v}_1^{\,\prime}$ and $\vec{v}_2^{\,\prime}$ as functions of   the
pre-collision  velocities,
\begin{equation}
  \label{eq:eps}
  \vec{v}_{1/2}^{\,\prime} = \vec{v}_{1/2} \mp  \frac{1+\varepsilon}{2}
  \left(\vec{v}_{12} \cdot \vec{e} \, \right)\vec{e}\,
\end{equation}
with $\vec{v}_{12}\equiv\vec{v}_1-\vec{v}_2$ and the unit vector  $\vec{e}
\equiv\left(\vec{r}_1-\vec{r}_2\right)/\left|\vec{r}_1-\vec{r}_2\right|$ at
the moment of the collision. 

In the HCS, the velocity distribution $f\left(\vec{v},\tau\right)$ (where the time
$\tau$ is measured in the average number
of collisions per particle) can be reduced to a time-independent
distribution function $\tilde{f} \left(\vec{c}\, \right)$ by the
transformation
\begin{equation}
\label{eq:defScalf}
f\left(\vec{v},\tau\right)=\frac{n}{v_T^3(\tau)} \tilde{f}\left(\vec{c}\,\right)\,,
\qquad\qquad \vec{c} \equiv \frac{\vec{v}}{v_T(\tau)} \,,
\end{equation}
with Haff's law \cite{Haff:83} for the temperature evolution,
\begin{equation}
\label{eq:dTdt}
\frac{dT}{d\tau}= -2\gamma T\,,~~\mbox{i.e.}\,~~~T(\tau)=T(0)\exp(-2\gamma\tau)\,.
\end{equation}

The main part of the distribution function, $c\sim 1$, can be described
with good accuracy by a second order Sonine polynomials expansion
around the Maxwell distribution $\phi(c)\equiv\pi^{-3/2}\exp(-c^2)$:
\begin{equation}
\label{eq:Sonexp}
\tilde{f}(c)=  \phi(c) \left[ 1 + a_1 S_1\left(c^2\right)+ a_2 S_2\left(c^2\right)+\dots \right]  \,.
\end{equation}
It can be shown that $a_1=0$, therefore, the leading deviations from the
Maxwell distribution are due to the second Sonine polynomial $S_2(c^2)
=c^4/2 - 5c^2/2 +15/8$ and the respective coefficient $a_2$
\cite{GoldshteinShapiro1:1995,NoijeErnst:1998,BrilliantovPoeschelStability:2000}:
\begin{equation}
\label{eq:a2_NE}
a_2 =\frac{16(1-\varepsilon) (1-2 \varepsilon^2)}{ 81 -17 \varepsilon +30 \varepsilon^2(1- \varepsilon)} \, .
\end{equation}
The good agreement of Eq.~\eqref{eq:Sonexp} with simulation data in
the region $c\sim 1$ can be seen in Fig. \ref{fig:distribution} (top).

For $c \gg 1$, Eq.~\eqref{eq:Sonexp} fails to represent the velocity
distribution function due to its  different functional form. It has
been shown that for particles with high velocities the distribution
function develops an exponential tail
\cite{EsipovPoeschel:1995,NoijeErnst:1998}, 
\begin{equation}
\label{eq:tail}
\tilde{f}(c) =  B e^{-bc} \, ;~
b=\frac{3\pi}{\mu_2}\,;~
\mu_2=\sqrt{2\pi}(1-\varepsilon^2)\left[1+\frac{3}{16}a_2\right]
\end{equation}
which is illustrated in the bottom part of
Fig.~\ref{fig:distribution}; here $\mu_2$ is the second moment of the
collision integral, e.g.  \cite{BrilliantovPoeschelOUP}.

The overpopulation of the tail is a rather general feature of granular
gases: After theoretically predicted \cite{EsipovPoeschel:1995}, it was found for gases in the HCS
also numerically \cite{BreyCuberoRuizMontero:1999,HuthmannOrzaBrito:2000} as well as for 
driven gases and was also detected experimentally, e.g.
\cite{Losert:1999,RouyerMenon:2000,OlafsenUrbach:1998,Kudrollietal:1997,SwinneyPRE:2004}.
In spite of its obvious importance, still it lacks a theoretical
description which allows to quantify its impact on granular gas
properties. Indeed, neither the numerical prefactor $B$ in
Eq.~\eqref{eq:tail} is known, nor the threshold velocity above which
the tail is overpopulated. In the present letter we address this
problem numerically and analytically. We develop an approach to
quantify the high-energy tail and estimate its impact on gas
properties. The impact of the tail on the cooling rate is studied in
detail.

We perform {\em Direct Simulation Monte Carlo} (DSMC)
\cite{Bird:1994,MontaneroSantos:1996,Puglisi:1999} of $N=10^8$
granular particles. DSMC is particularly suited to simulate large
systems over long time in the HCS, that is, to suppress spatial
correlations which give rise to vortices \cite{BritoErnst:1998} and
clusters \cite{GoldhirschZanetti:1993}. We started at $T(0)=1$ and
simulated until the particle velocities approached the double
precision number representation, i.e., until $T\approx 10^{-23}$. For
$\varepsilon=0.9$ this corresponds to a total of $5\times 10^{10}$
collisions or 1,000 collisions per particle. 
Neglecting the first $2\cdot 10^9$ collisions, after each $10^8$
collisions (2 collisions per particle) we recorded a snapshot of the scaled velocities, $c$.
The distribution function, $\tilde{f}(c)$, was then obtained by binning of up to 100 of 
such snapshots in 100 intervals. 

The numerical values of the constants $B$ and $b$ were then determined
by performing a least-mean-square fit of the linear function
$B^{\prime} -bc$, where $B^{\prime} =\log B$. To find the numerical
value of the threshold velocity where the tail starts, $c_*$, (see the
discussion below) we discriminate between two cases. For the first
case, where the functions $B\exp(-b c)$ and $\tilde{f}(c)$ according
to the Sonine expansion, Eqs. \eqref{eq:Sonexp}, \eqref{eq:a2_NE}
intersect, we determine $c_*$ as the mean of the two intersection
points; these were very close to each other in all our
simulations. For the second case, where the functions do not
intersect, $c_*$ was determined as the scaled velocity that minimizes
the distance between both functions.  Such simulations were performed
for $\varepsilon=0.1, 0.2, \dots, 0.9$.

\begin{figure}[tbp] 
  \centerline{\includegraphics[width=7cm,clip]{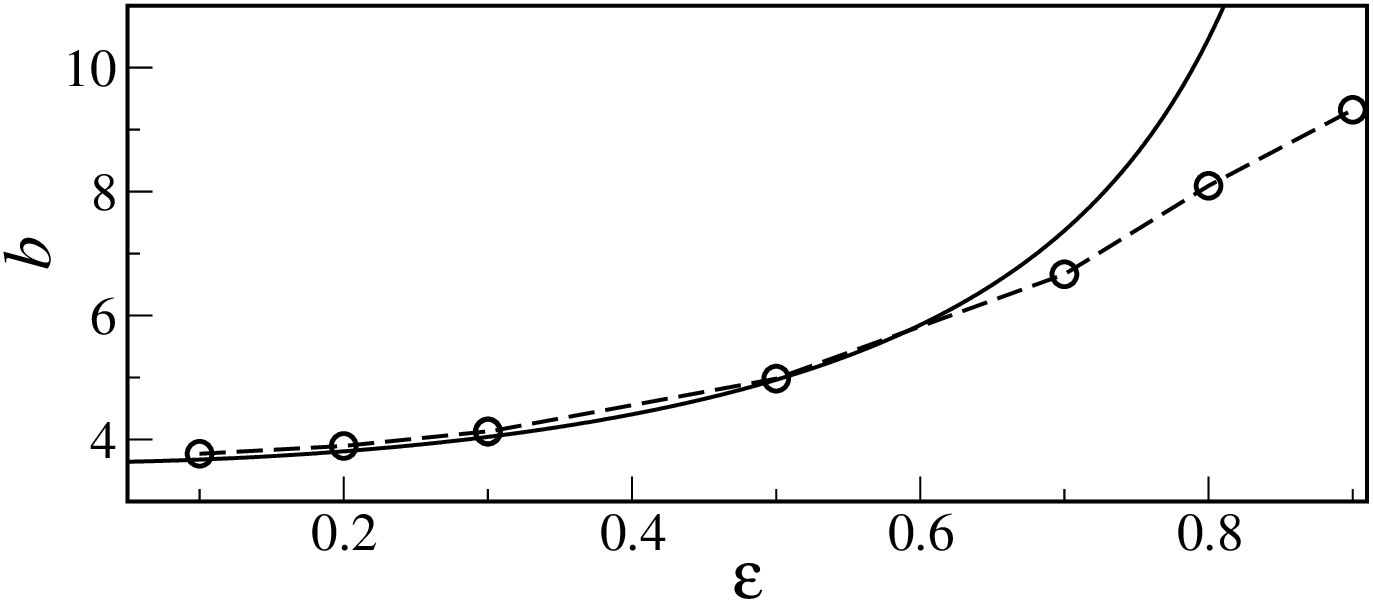}}
  \caption{The slope $b$ of the exponential tail. For
    $\varepsilon\lesssim 0.7$ we find good agreement of the DSMC
    results with the theoretical expression given in Eq.
    \eqref{eq:tail} (full line). For smaller dissipation the data
    deviate (see text).}
\label{fig:b} 
\end{figure}

In Fig.~\ref{fig:b} the numerical results for the slope $b$ are
compared with theoretical prediction, Eq.~\eqref{eq:tail}.  For
$\varepsilon\lesssim 0.7$ we find good agreement, whereas for larger
$\varepsilon$ the data deviate. This can be understood from the
theoretical argument used to derive $\tilde{f}(c) \propto\exp(- b\,
c)$ \cite{EsipovPoeschel:1995,GoldhirschetalNLP:2003}: For $c \to
\infty$ the gain term of the Boltzmann equation may be neglected as
compared with the loss term. The larger the restitution coefficient
$\varepsilon$, the later the tail starts.  Contrary, for molecular
gases where $\varepsilon=1$, the gain and loss terms balance each
other for all velocities leading to the Maxwell distribution,
$\tilde{f}(c)\propto \exp(-c^2)$. The deviation for $\varepsilon
\gtrsim 0.7$ originates from the fact that the gain term cannot be
neglected for large $\varepsilon$ for the accessible interval of
velocities.

\begin{figure}[htbp]
\centerline{\includegraphics[width=7.5cm,clip]{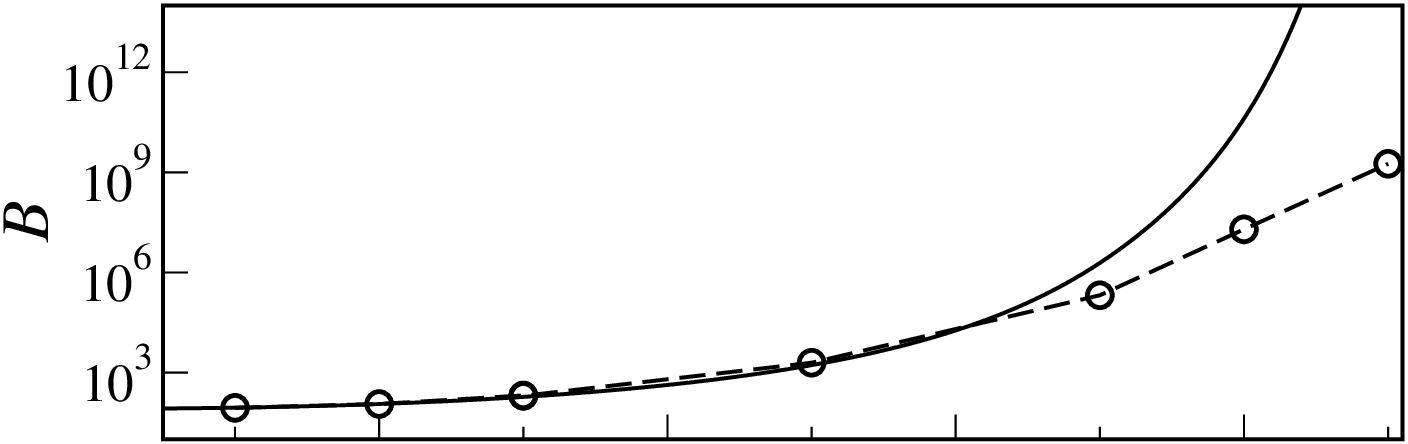}}
\centerline{\includegraphics[width=7.5cm,clip]{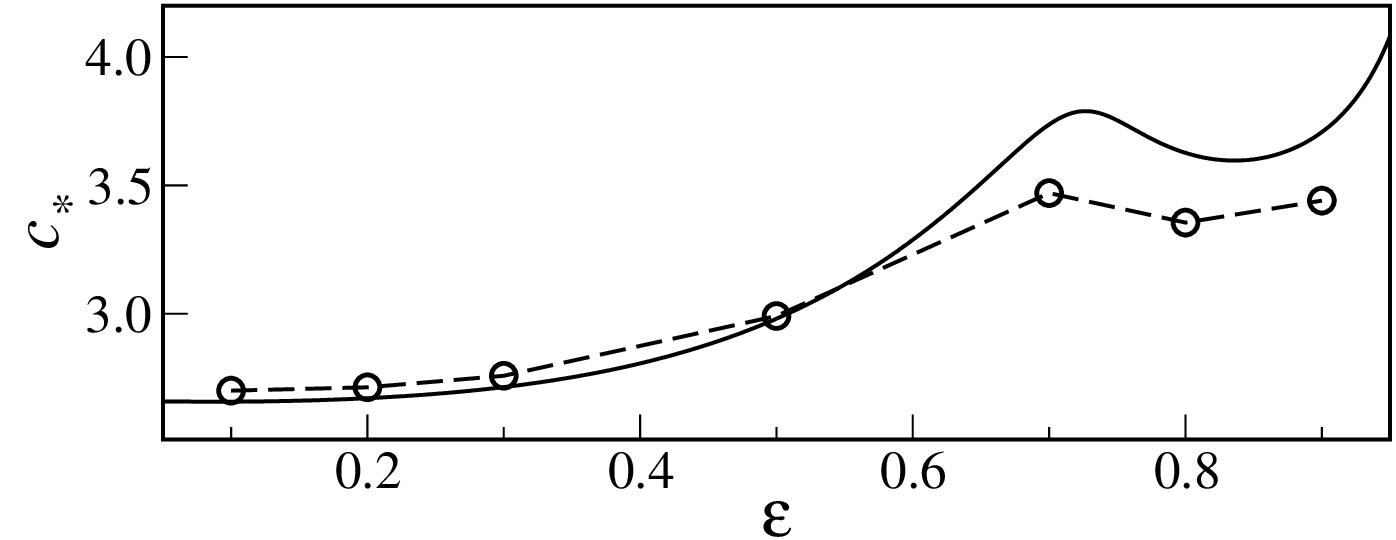}}
\caption{The threshold velocity $c_*$ and the parameter $B$ as
  obtained from DSMC (symbols) together with the solution of Eqs.
  \eqref{eq:eq_for_cstar} and \eqref{eq:def_B}. Again, for
  $\varepsilon\lesssim 0.7$ we find good agreement.}
\label{fig:B}
\end{figure}

To obtain the the threshold velocity  $c_*$ analytically we   assume that
$f(c)$ may be  sufficiently well described by a
combination of Eq. \eqref{eq:Sonexp}, valid for $c\sim 1$ and
Eq.~\eqref{eq:tail} for the tail,
\begin{equation}
\label{eq:ftentat}
\tilde{f}(c) = A c^2 e^{-c^2} \left[ 1 + a_2 S_2(c^2) \right] \theta (c_*-c)
  + B c^2 e^{-bc} \theta (c-c_*) \,,
\end{equation}
with the Heaviside function $\Theta(x)$, i.e., we disregard the
transient region \cite{GoldhirschetalNLP:2003} between the
near-Maxwellian and exponential parts of the distribution. Using DSMC
we checked the Ansatz \eqref{eq:ftentat}, whose eligibility is
illustrated in Fig.~\ref{fig:distribution} for $\varepsilon=0.3$.  The
unknown parameters $A$, $B$ and the threshold velocity $c_*$ may be
found from the normalization condition and continuity of the
distribution function itself and its first derivative 
\begin{equation}
\label{eq:der_and_func}
\tilde{f}(c_*\!+0) = \tilde{f}(c_*\!-0)  \, , \qquad  
\tilde{f}^{\prime}(c_*\!+0) = \tilde{f}^{\prime}(c_*\!-0) \, ,  
\end{equation}
where
$\tilde{f}^{\prime}= d \tilde{f}/dc$.  From normalization follows then
\begin{eqnarray}
\label{eq:eq_for_cstar}
c_*&=&\frac{b}{2} +\frac{a_2 \left( 2 c_*^3 -5 c_* \right)}{2 \left(1+a_2 S_2(c_*^2)\right)}\\
\label{eq:def_A}
A^{-1}&=& \frac{k(c_*)}{b^3}  \left( 2 + b c_* (2 + b c_* ) \right) e^{-bc_*}
+ \frac{\sqrt{ \pi} }{4} {\rm     Erf} (c_*)  \nonumber \\
&& ~~~~~~~~~~~~ -\frac18 c_* (4+a_2 c_*^2(2 c_*^2 -5)) e^{-c_*^2}   \\
\label{eq:def_B}
B &=& A k(c_*)
\end{eqnarray}
with
\begin{equation}
\label{eq:def_k}
k(c_*) \equiv  e^{-c_*^2 + b c_*} \left( 1 + a_2 S_2 ( c_*^2 ) \right) \, .
\end{equation}
Solving the fifth order equation \eqref{eq:eq_for_cstar} numerically
for $c_*$, we obtain $A$, $k$ and finally $B$, Fig. \ref{fig:B}.

The very good agreement between simulations and the theoretical
predictions for the coefficients $A(\varepsilon)$, $B(\varepsilon)$
and the transition velocity $c_*(\varepsilon)$ manifests the adequacy
of the Ansatz \eqref{eq:ftentat}.  The deviations for
$\varepsilon\gtrsim 0.7$ occur already for the slope $b$ of the
exponential tail, Fig.~\ref{fig:b}, which is not related to the
Ansatz. It may be explained similarly as the deviation of $b$ from its
theoretical value: For large $\varepsilon\gtrsim 0.7$ the system of
$N=10^8$ particles is not large enough to develop a well-detectable
exponential tail.

Knowing the distribution function, Eq.~\eqref{eq:ftentat} and its
parameters $A$, $B$ and $c_*$ as functions of $\varepsilon$, we can
quantify the impact of the exponential tail on kinetic quantities. In
this letter we focus on the temperature decay rate $\gamma$; for the
diffusion coefficient, other transport coefficients as well as for
technical details we refer to \cite{laterpaper}.

The standard analysis (e.g. \cite{NoijeErnst:1998,BrilliantovPoeschelOUP})
yields for the temperature decay rate, when the stationary velocity
distribution is achieved,
\begin{equation}
\label{eq:gam_gen}
\gamma=(1-\varepsilon^2)\frac{J_2}{12J_0} \, ,
\end{equation}
where 
\begin{equation}
\label{eq:kin_int}
J_k \!= \! \int \! d \vec{c}_1  \! d \vec{c}_2   \int \! d \vec{e} \,
 \Theta(-\vec{c}_{12} \cdot \vec{e} \, ) |\vec{c}_{12} \cdot \vec{e}\, |^{k+1} 
 \tilde{f}(c_1) \tilde{f}(c_2)  \, .
\end{equation}
Disregarding the exponential tail, the energy decay rate reduces to
\cite{NoijeErnst:1998}
\begin{equation}
  \label{eq:gamma_0}
  \gamma_0(\varepsilon)=\frac{1-\varepsilon^2}{6}\frac{1+(3/16)
  \,a_2(\varepsilon)}{1-(1/16) \,a_2(\varepsilon)}
\end{equation}
\begin{figure}[htbp]
\centerline{\includegraphics[width=8cm,clip]{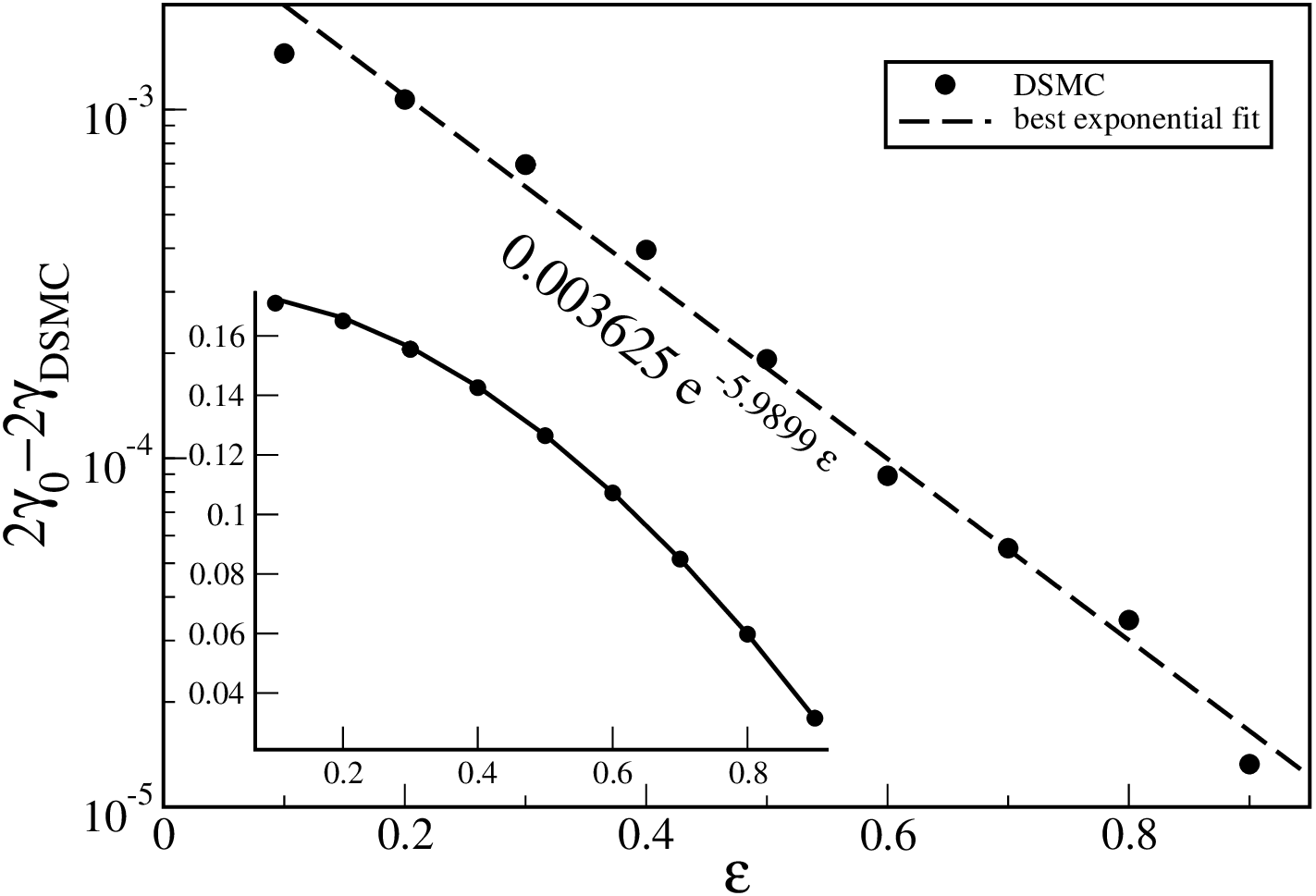}}
\caption{Inset: The temperature relaxation coefficient
  $\gamma(\varepsilon)$ over $\varepsilon$ as obtained by DSMC,
  $\gamma_{\rm DSMC}$, (points) and due to Eq. \eqref{eq:gamma_0},
  $\gamma_0$. The logarithmic plot shows $(\gamma_0-\gamma_{\rm
    DSMC})$, that is, the influence of the exponential tail on the
  cooling coefficient. The dashed line shows the best exponential
  fit.}
\label{fig:gamma}
\end{figure}
We applied DSMC of $10^8$ particles for different $\varepsilon$ and
recorded the temperature $T_{\rm DSMC}(\tau)$.  Then $\gamma_{\rm
  DSMC}$ was determined by fitting $T_{\rm DSMC}(\tau)$ for $\tau\gg
1$ to its asymptotic law, $T_{\rm DSMC}\propto\exp(-2\,\gamma_{\rm
  DSMC}\,\tau)$, Eq.~\eqref{eq:dTdt}. Figure \ref{fig:gamma} (inset)
shows $\gamma_{\rm DSMC}(\varepsilon)$ (points) together with the
analytical result, $\gamma_0(\varepsilon)$, Eq.~\eqref{eq:gamma_0}
(line). In this representation we hardly see any discrepancy between
theoretical and numerical data. The difference between these curves
(main part of Fig. \ref{fig:gamma}) which quantifies the impact of the
overpopulated tail on the cooling rate reveals, however, a clear
dependence on $\varepsilon$. The scaling law, $\gamma_0-\gamma_{\rm
  DSMC}\propto \exp(-6\varepsilon)$, shown here only as a numerical
result is, however, difficult to confirm analytically: In spite of the simple
functional form of $\tilde{f}(c)$, Eq. \eqref{eq:ftentat}, an accurate
analytical expression for the cooling coefficient may be obtained only in the limit
$\varepsilon  \lesssim 1$ \cite{laterpaper}. As it follows from the discussion
below, the studied system of $N=10^8$ particles is not sufficiently large to
analyze this limit by numerical simulations. 

So far, the discussion refers to the state when the velocity
distribution $\tilde{f}(c)$ has relaxed to its stationary form. Now we
ask the question, how fast does this happen?  To address this problem
we initialize the particle velocities according to a Maxwell
distribution at $T(0)=1$ and investigate the decay of temperature as a
function of the average number of collisions $\tau$. Asymptotically,
i.e., when the gas has adopted its asymptotic distribution, the
temperature evolves according to Haff's law, Eq.~\eqref{eq:dTdt}.
Thus, the time lag which is needed for a system to reach Haff's
evolution corresponds to the relaxation time of the distribution
function to achieve its stationary form $\tilde{f}(c)$.

\begin{figure}[htbp]
\centerline{\includegraphics[width=7.5cm,clip]{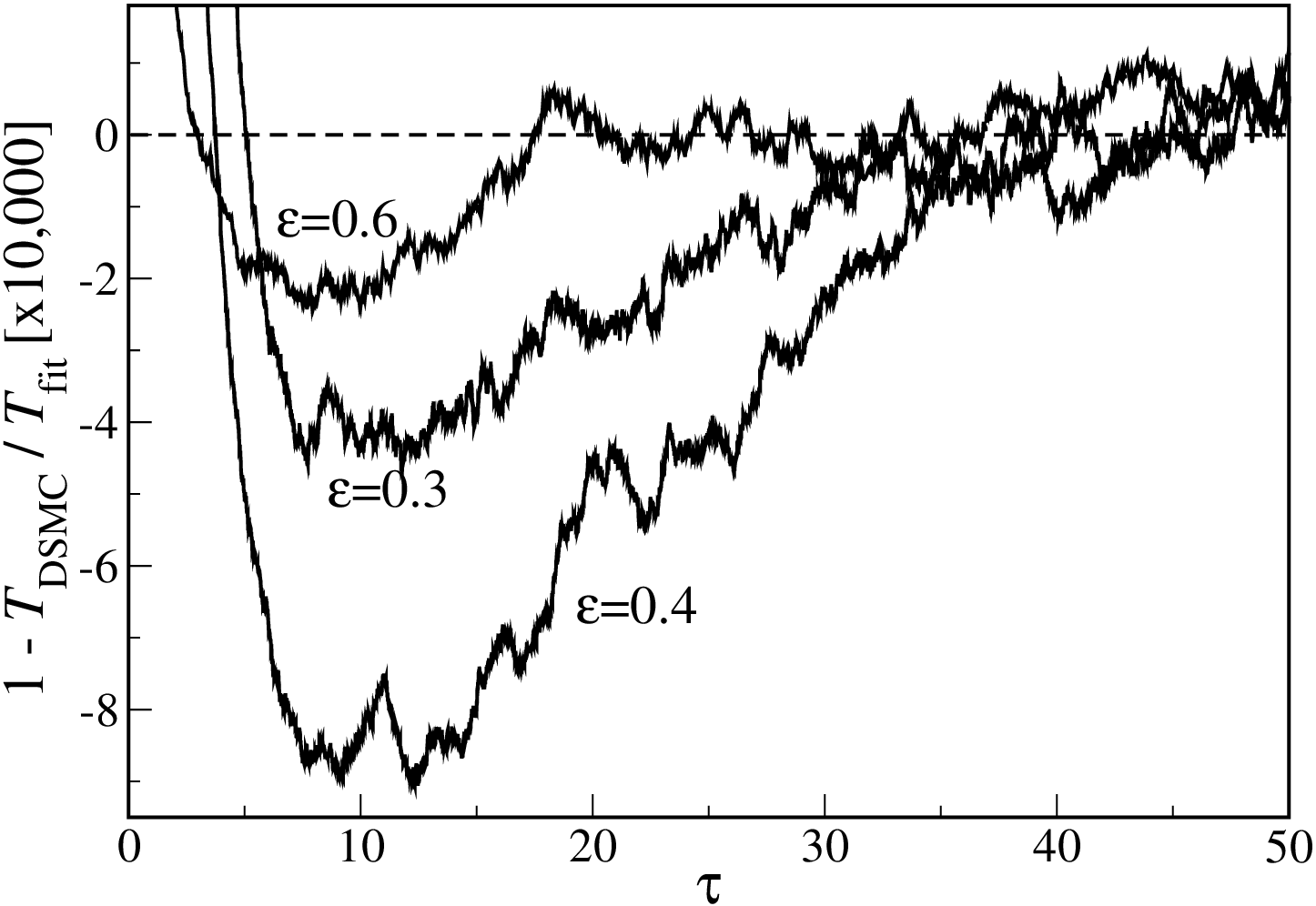}}
\caption{Relaxation of the temperature decay to Haff's law for $N=10^8$ and
  $\varepsilon=0.3, 0.4, 0.6$, characterizing the relaxation of the
  distribution function to its stationary  form.}
\label{fig:Trelax}
\end{figure}

Using the coefficient $\gamma_{\rm DSMC}$ described above, we define
the temperature $T_{\rm fit}(\tau) \propto \exp(-2\gamma_{\rm
  DSMC} \, \tau)$. By definition, for $\tau \gg 1$ we have $T_{\rm
  DSMC}\approx T_{\rm fit}$ since $\gamma_{\rm DSMC}$ was determined as
the best exponential fit to $T_{\rm DSMC}(\tau)$ for $\tau \gg 1$.
Therefore, the quantity $1-T_{\rm DSMC}(\tau)/T_{\rm fit}(\tau)$
characterizes the relaxation of the distribution function to its
stationary form. Figure \ref{fig:Trelax} shows the relaxation for
different values of the coefficient of restitution. We note that
depending on $\varepsilon$, the relaxation to the level of ``natural''
fluctuations in the system takes approximately 20 to
30 collisions per particle. This slow relaxation is very different
from that in a molecular gas, where it takes very few (3-5) collision
per particle to develop the Maxwell distribution.
Similar slow relaxation of the  high-energy tail of the velocity distribution
was reported for a gas of elastic hard spheres \cite{TailHardSpheres:1990}. The
relaxation mechanism for the case of granular  gases is, however, completely
different and depends on  $\varepsilon$: While the tail in elastic gases is
fed by the gain term of the Boltzmann equation, for the dissipative gases this
term is negligible. The formation of the tail for the latter case occurs
exclusively due to permanent cooling of the gas, so that the scaled velocity of a
particle,   $\vec{c}\equiv \vec{v}/v_{T}$, increases due to decaying
$v_T$. By this mechanism particles  enter the tail  and keep it overpopulated. 

The relaxation time, i.e., the time of formation of the exponential tail, may
be quantified.  The  relaxation to the stationary distribution $\tilde{f}(c)$
is  described by \cite{BrilliantovPoeschelOUP,BrilliantovPoeschel:2000visc}

\begin{equation}
\label{eq:kin_eq_for_fc}
\frac{\mu_2}{3} \left( 3 + c \frac{\partial}{\partial c } 
\right)  \tilde{f}(c,\tau )+ J_0 \frac{\partial}{\partial \tau  } \tilde{f}(c,\tau) = \tilde{I}
\left( \tilde{f},  \tilde{f} \right)
\end{equation}
where $\tilde{I} ( \tilde{f}, \tilde{f} )$ is the reduced
collision integral (e.g.  \cite{BrilliantovPoeschelOUP}) and $J_0$ is defined
in Eq.~\eqref{eq:kin_int}. Neglecting the incoming term  for $c \gg 1$ 
 \cite{EsipovPoeschel:1995,NoijeErnst:1998,MontaneroSantosTailsGG:1999}, the
collision integral may be approximated by 
\begin{equation}
\label{eq:coll_int}
\tilde{I} \left( \tilde{f} , \tilde{f} \right) \approx - \pi c \tilde{f}(c) \, , 
\qquad \qquad
c \gg 1 \, . 
\end{equation}
Using the Ansatz $\tilde{f}(c,\tau ) =B \exp \left[ -w(\tau ) c
\right] $, we recast Eq. \eqref{eq:kin_eq_for_fc} into
\begin{equation}
\label{eq:dwdt}
\frac{d w}{d \tau} + \frac{\mu_2}{3J_0}  w = \frac{\pi}{J_0}  \,  
\qquad \qquad
c \gg 1 \, 
\end{equation}
with the solution
\begin{equation}
\label{eq:wt_result}
w( \tau ) = b + (1-b) \exp\left[-\tau/ \tau_0(\varepsilon) \right]  \, ,
\end{equation}
where $b= 3 \pi / \mu_2$ coincides with Eq.~\eqref{eq:tail} and $
\tau_0^{-1}(\varepsilon) = \mu_2/3 J_0$. Neglecting $a_2$, which
characterizes small deformations of the main part of the distribution
with respect to the Maxwellian, and the contribution from the tail, we
obtain $J_0 =2 \sqrt{2 \pi}$ and hence 
\begin{equation}
\tau_0^{-1}(\varepsilon) = \frac{1-\varepsilon^2}{6} 
\label{eq:tau0}
\end{equation}
(see \cite{laterpaper} for details). For
$\varepsilon =0.4$ we obtain the relaxation time, $\tau_0 \approx
7.1$. 

Let us compare the theoretical prediction, Eq. \eqref{eq:tau0}, with
the numerical results: The relaxation of $(1-T_{\rm DSMC}/T_{\rm
fit})$ as plotted in Fig.  \ref{fig:Trelax} reveals two
stages. First, for $\tau\lesssim 10$, we observe
relaxation of the main part of the velocity distribution, where
$c\approx 1$. The initial Maxwell distribution relaxes here to the
distribution given by the Sonine expansion,
Eq. \eqref{eq:Sonexp}. During the second stage the overpopulated tail
is formed; the plotted quantity decreases for $\varepsilon =0.4$ by a
factor of $10$ in the time span $\Delta \tau = 25$ ranging from
$\tau=10$ to $\tau=35$. This leads to a numerical relaxation time
$\tau_0=25/\log(10) \approx 10.8$, in agreement with the above
theoretical estimates.

The theory also predicts that the relaxation time increases with increasing
$\varepsilon $. While this tendency is confirmed for $\varepsilon =0.3$ and
$\varepsilon =0.4$, it is seemingly violated for $\varepsilon =0.6$, see Fig.
\ref{fig:Trelax}. We argue however that this  is, presumably, a finite size
effect, which may be understood as follows. According to the mechanism of the
tail formation, discussed above, the gain term of the collision integral does
not contribute to the tail. Instead, particles enter the tail due to increase
of the scaled velocity,  $\vec{c}\equiv \vec{v}/v_{T}$, when the thermal
velocity $v_{T}$ decreases along with temperature $T$.  The
temperature  decay and hence the formation of the tail is
slower  for larger $\varepsilon$, that is,  the relaxation time $\tau_0$ is
larger,  Eq. \ref{eq:tau0}. 

On the other hand, the total number of particles in
the tail moving at velocities $c>c*$,  decreases with increasing
$\varepsilon$.  Correspondingly, the deviation of the distribution function from its
steady state  $\tilde{f}(c)$,  quantified here by $(1-T_{\rm DSMC}/T_{\rm fit})$, becomes
smaller for smaller dissipation. Consequently, the relaxation of this quantity may
be traced only as long as it exceeds the level of natural
fluctuations. The smaller the system,  the larger is the impact of the fluctuations.
Therefore, if  the number  of particles
in the tail is not sufficiently large, the value of $(1-T_{\rm DSMC}/T_{\rm
  fit})$ drops quickly below the fluctuation level, making an accurate
numerical estimate of the relaxation time impossible. This is the case for $\varepsilon
=0.6$ in Fig. \ref{fig:Trelax} where a seemingly fast relaxation is
observed due to large fluctuations. Hence, we conclude that the observed
relaxation curves do not contradict the predictions of the theory. They
indicate, however, that the size of the system  of $N=10^8$ particles is not
sufficient to study the relaxation of the distribution function for $\varepsilon
=0.6$ or larger. For a {\em very} large system we expect increase of the
relaxation time $\tau_0$ with increasing coefficient of restitution
$\varepsilon$ in agreement with the theoretical analysis.

In summary, we investigated the velocity distribution function of a
granular gas and the impact of its overpopulated high-energy tail on
the cooling coefficient, which is the main characteristics of a
granular gas in the HCS. We proposed a unified functional form of the
distribution function which comprises its main part ($v/v_T \equiv c
\sim 1 $) whose deviation from the Maxwell distribution is described
by the second-order Sonine expansion, and the overpopulated tail which
decays exponentially.  We derived $\varepsilon$-dependent coefficients
of the proposed Ansatz along with the scaled velocity $c_*$, which
separates the main part of the velocity distribution ($c\approx 1$)
and the tail part ($c\gg 1$). For $\varepsilon\lesssim 0.7$ the
analytical results agree well with large scale DSMC of $10^8$
particles, while the deviations for $\varepsilon \gtrsim 0.7$ may be
attributed to finite size effects.

We analyzed the impact of the overpopulated high-energy tail
on the cooling rate $\gamma$ which is the main hydrodynamic
coefficient of granular gases in the HCS. We found 
systematic deviations from the theoretical expression which neglects 
the exponential tail. These deviations grow with
increasing dissipation (decreasing $\varepsilon$) as
$\exp(-6\varepsilon)$, due to enhanced contributions from the tail.

Finally, we observed and explained theoretically the extraordinary
slow (as compared with molecular gases) relaxation of the velocity
distribution to its asymptotic stationary form. It takes about $\sim
20-30$ collisions per particle and may be understood from the
mechanism of the tail formation.

\acknowledgments{
This research was supported by a Grant from the G.I.F., the German-Israeli Foundation for Scientific Research and Development.}

\end{document}